\documentclass[preprint]{aastex}
\usepackage{amsmath}
\usepackage{natbib}

\begin{document}

\bibliographystyle{apj}

\title{The Atmospheric Signatures of Super-Earths: How to Distinguish Between Hydrogen-Rich and Hydrogen-Poor Atmospheres}

\author{Eliza Miller-Ricci}

\affil{Harvard-Smithsonian Center for Astrophysics, 60 Garden St. Cambridge,
        MA 02138}

\email{emillerricci@cfa.harvard.edu}

\author{Sara Seager}

\affil{Department of Earth, Atmospheric, and Planetary Sciences, Department
       of Physics, Massachusetts Institute of Technology, 54-1626, 77
       Massachusetts Ave., Cambridge, MA 02139}

\author{Dimitar Sasselov}

\affil{Harvard-Smithsonian Center for Astrophysics, 60 Garden St. Cambridge,
        MA 02138}

\begin{abstract}

Extrasolar super-Earths (1-10 M$_{\earth}$) are likely to exist with a wide
range of atmospheres. Some super-Earths may be able to retain massive 
hydrogen-rich atmospheres.  Others might never accumulate hydrogen or 
experience significant escape of lightweight elements, resulting in atmospheres
more like those of the terrestrial planets in our Solar System.  We examine 
how an observer could differentiate between hydrogen-rich and hydrogen-poor
atmospheres by modeling super-Earth
emission and transmission spectra, and we find that 
discrimination is possible by observing the transmission spectrum alone.  An 
Earth-like atmosphere, composed of mostly heavy elements and molecules, will 
have a very weak transmission signal due to its small atmospheric scale height
(since the scale height is inversely proportional to molecular weight).  
On the other hand, a large hydrogen-rich atmosphere reveals a relatively large 
transmission signal.  The super Earth emission spectrum can 
additionally contrain the atmospheric composition and temperature structure. 
Super-Earths with massive hydrogen atmospheres will reveal strong 
spectral features due to water, whereas those that have lost most of their
hydrogen (and have no liquid ocean) will be marked by CO$_2$ features and a 
lack of H$_2$O.  
We apply our study specifically to the low-mass planet orbiting an M star,
Gl 581c ($M sin i$ =  5 M$_{\earth}$), although our conclusions are relevant
for super-Earths in general.  The ability to distinguish hydrogen-rich 
atmospheres might be essential for interpreting mass and radius observations 
of planets in the transition between rocky super-Earths and Neptune-like 
planets.

\end{abstract}


\section{Introduction}

The search for extrasolar planets has recently resulted in the discovery of
a new class of planets, super-Earths, with masses between about 1 and 10 Earth
masses (M$_{\earth}$) \citep{riv05, bea06, udr07}.  
Since no such planets exist in our Solar System, the
bulk composition of these planets and of their atmospheres remains largely 
unknown, although some theoretical work on the subject has been presented.  
In contrast to rocky planets in our 
Solar System, super-Earths are predicted to have large surface gravities, 
- upwards of 25 m/s$^2$ for a 5 M$_{earth}$ planet - 
\citep{val06, val07, for07, sea07, sot07} and some super-Earths may 
therefore be able to retain massive H-rich 
atmospheres.  Others will bear a closer resemblence to Earth, with atmospheres
depleted in hydrogen and composed of predominantly heavier molecules. 
A third interesting case is that some super-Earths may lie in 
between these two extremes, with moderate levels of atmospheric hydrogen due to
either incomplete escape of hydrogen over the planet's lifetime and/or from
outgassing of a significant secondary H$_2$ atmosphere.  
The question arises as to whether these cases will differ in their gross 
observable properties, and how. In the interest of 
discovering habitable planets and learning more about atmospheric evolution on
super-Earths, it will be necessary to find a method for discriminating between
the possible types of atmospheres.  This issue has led us to 
model the observational signature for the various possible atmospheres on the 
low-mass planet, Gl 581c \citep{udr07}.  

The bulk composition of planets in the mass range between Earth and Neptune 
carries important clues to understanding planet formation.  Even the simple 
statistical distribution of very water-rich versus dry super-Earths in an 
observed sample would be sufficient \citep{val07}. Unfortunately, with planet 
mass and radius alone the internal structure models have degeneracies, which 
become almost untractable when a hidrogen-rich envelope covers the planet
\citep{ada07}. Our study in this paper points to a potential solution to 
breaking this degeneracy, by distinguishing a hydrogen-rich atmosphere 
spectroscopically. 

A number of observations of transiting hot Jupiters have already allowed for
constraints to be placed on the nature of their atmospheres.  IR observations
of the secondary eclipse (when the planet passes behind its host star) have
resulted in measurements of brightness temperatures \citep{dem05, cha05, 
dem06, har07, dem07, demo07, knu08}
as well as emission spectra \citep{gri07, ric07, swa07, cha08} for several hot 
Jupiters.  Optical transmission spectra for HD 209458b have led to the 
discovery of several species in the planet's atmosphere including sodium 
\citep{cha02}, hydrogen \citep{vid03, ehr08}, and tentative detections of 
oxygen and carbon \citep{vid04}.  
Recent observations of the IR transmission spectrum of HD 189733b have
led to the first discoveries of water \citep{tin07, bea08} and methane 
\citep{swa08} in the atmosphere of an extrasolar planet.  Additionally, 
attempted observations of the optical secondary eclipse of HD 209458b using 
the {\it MOST} space telescope have allowed for strong limits to be placed on 
the planet's optical albedo \citep{row06, row07}.  Since super-Earths are 
smaller than hot Jupiters,
these types of measurements will prove more challenging for this new class of
planets \citep{sea08}.  However, it will most likely be these same types of 
observations 
that will eventually aid in characterizing the  atmospheres of low-mass 
planets.

The discovery of the low-mass planet Gl 581c ($M sin i =  5 M_{\earth}$) has 
generated a large amount of interest in the literature, and a number of 
authors have already explored the question of habitability for this planet 
after an initial claim by \citet{udr07} that Gl 581c may reside inside its 
star's habitable zone.  It now appears that the planet is probably too hot to 
host liquid water unless its surface is mostly obscured by 
highly reflective clouds \citep{sel07}, or the planet is tidally locked, with 
the night side receiving sufficient heat circulation from the day side of the 
planet \citep{chy07}.  To further confirm this conclusion, 
\citet{von07, von07b} have examined the possibility of 
sustained photosynthetic life on Gl 581c and have ruled this out on the grounds
that the planet cannot keep sufficient CO$_2$ in its
atmosphere for photosynthesis to occur, while at the same time retaining a
surface temperature cool enough for liquid water.  However, in terms of
climate stability,  \citet{beu07} have modeled the dynamic evolution 
of the Gl 581 system.  They find that the orbit (and therefore the climate) 
of Gl 581c should be secularly stable over 10$^8$ yrs, leaving open a 
possibility that life of a more exotic variety could still develop on this 
planet.   

In this paper, we address a different question relating to the nature of
Gl 581c's atmosphere.  What observations are needed to 
constrain the atmospheric 
composition and overall hydrogen content for Gl 581c or similar super-Earths?  
To this end, we 
investigate several scenarios for the composition of the super-Earth 
atmosphere ranging from hydrogen-rich to hydrogen-poor. 
The cases that we explore are (i) a massive hydrogen-rich (reducing) 
atmosphere 
obtained through accretion of nebular gas, (ii) a mildly reducing atmosphere
that has lost most but not all of its hydrogen due to atmospheric escape
processes, and (iii) an oxidizing CO$_2$-rich atmosphere similar in composition
to that of Venus that contains essentially no hydrogen.  From the modeled 
spectra, we determine how to observationally
differentiate between the various scenarios.  We describe our model atmosphere
in \S~\ref{model} and our three scenarios for atmospheric composition in 
\S~\ref{comp}.  In \S~\ref{results} we reveal the modeled emission and
transmission spectra for Gl 581c, and we determine how hydrogen content can
be established through observations.  Finally, we summarize our 
results and discuss their implications in \S~\ref{discussion}.

\section{Super Earth Atmosphere Model \label{model}}

\subsection{Opacities}

We determine molecular opacities following the procedure outlined in
\citet{sea00} for wavelengths ranging from 0.1 to 100 $\mu$m.
We include the dominant sources of molecular line opacity in the IR: 
CH$_4$, CO, NH$_3$, 
\citep[][and references therein]{fre07}, H$_2$O \citep{fre07, par97}, CO$_2$, 
O$_2$, and O$_3$ \citep{rot05}.  CO$_2$ - CO$_2$ collision induced opacities
are also included \citep{bro91}, which we find to have an effect on the 
near-IR portion of some of the super-Earth spectra.  
We have not included opacities of any 
condensed species including the solid and liquid forms of water, since we find 
that none of 
our atmospheres enter into the temperature and pressure regime of water clouds.

\subsection{Chemistry \label{chemistry}}

Given the atomic makeup of our super-Earth atmosphere, we determine 
molecular abundances in chemical equilibrium.  This is accomplished by 
minimizing Gibbs free energy following the method outlined in \citet{whi58} 
for 23 atoms and 172 gas phase molecules.  For the Gibbs
free energies of each molecule we use polynomial fits from \citet{sha90}, 
which we find to be a good match to data from the NIST JANAF thermochemical 
tables \citep{cha98} over our full temperature range of 
interest (100-1000 K).  We do not include a full treatment of condensed
species in our chemical model, although we do determine whether any of our 
model atmospheres cross the condensation curves of known cloud-forming 
species.  Specifically, we look for condensation of H$_2$O and stability of 
the H$_2$SO$_4$ condensate,
since both are known to form clouds on planets of the inner Solar System and
could potentially alter the T-P profile and planetary emission spectrum 
predicted by our models. 

We find that chemical equilibrium for a super-Earth atmosphere 
containing large quantities 
of hydrogen predicts high abundances of both ammonia and methane.  However,
these two species are known to be destroyed readily by UV photolysis in 
planetary atmospheres.  Determining NH$_3$ and CH$_4$ abundances in 
photochemical equilibrium would involve knowledge of the full
range of chemical reactions that take place in the super-Earth atmosphere 
and their reaction rates as a function of temperature.  Compiling such a 
model for the range of atmospheric chemistries that we consider for Gl 581c is 
beyond the scope of this paper.  Instead, we calculate photolysis lifetimes
for both NH$_3$ and CH$_4$, and we assume that return reactions replenishing 
these molecules occur slowly, which is generally correct in the temperature 
range that we consider.  Molecules are then determined to be photochemically 
unstable if their photolysis 
lifetimes, $1/J(z)$, are found to be less than 
the age of the planet (approximately 4.3 Gyr \citep{udr07}) throughout a 
significant portion of the atmosphere.  Since atmospheric overturning
timescales should be short - on the order of hours \citep{mas81} - mixing
will occur readily to regions where the molecules are destroyed by 
photolysis.  Here $J(z)$ is the photodissociation
rate at height $z$ in the atmosphere:
\begin{equation}
J(z) = \int \sigma_{\nu} F_{\nu} e^{-\tau_{\nu}} \,d\nu, \label{photochem}
\end{equation}
where $\sigma_{\nu}$ is the photodissociation cross-section at frequency $\nu$
and  $F_\nu$ is the incident solar flux expressed in photons 
cm$^{-2}$ s$^{-1}$ Hz$^{-1}$.  
The optical depth, $\tau_{\nu}$ is given by
\begin{equation}
\tau_{\nu}(z) = \int_{z}^{z_{max}} \kappa_{\nu} \,ds,
\end{equation}
where $\kappa$ is the opacity in units of cm$^{-1}$.  Here we have included 
contributions both from absorption of photons and from Rayleigh scattering, 
which can have a significant effect in shielding the atmosphere from UV 
penetration at the shortest wavelengths.  We obtain UV cross-sections for 
the major constituents of the atmosphere from \citet{yun99}.  The UV flux from
Gl 581 is taken from IUE observations as reported by \citet{buc07} and scaled 
to a distance of 0.073 A.~U.   

If methane and ammonia are found to be sufficiently short-lived, we then 
remove these species from our chemical model.  In 
practice this is accomplished by recalculating
chemical equilibrium without the presence of NH$_3$ and CH$_4$.  
Their atomic constituents are then reallocated to the most stable 
molecules - N$_2$ and H$_2$ in 
the case of ammonia, and H$_2$, CO, and CO$_2$ for methane.  The resulting 
chemical compositions for the Super-Earth atmospheres that we consider are 
presented in Section~\ref{comp}.

\subsection{Irradiated Temperature-Pressure Profile \label{t_p}}

We compute a 1-D radial radiative-convective temperature-pressure (T-P) 
profile for Gl 581c. 
We have taken a fairly basic approach toward generating the T-P profile in
that we assume a grey atmosphere in hydrostatic equilibrium.  However, we
point out that the 
major results we present later in this paper for transmission spectra 
are not strongly dependent on the exact T-P profile that is used.  

For the temperature 
structure, we make the assumption of an irradiated grey atmosphere as 
presented by \citet{che07} and \citet{han07}.  The planet's temperature 
profile as a function of optical depth $\tau$ is described by
\begin{equation}
T^{4} = \frac{3}{4} T_{eff}^{4} \left[{\tau + \frac{2}{3}}\right] + 
\mu_0 T_{eq}^{4} \left[1 + \frac{3}{2} \left(\frac{\mu_0}{\gamma}\right)^2 - 
\frac{3}{2} \left(\frac{\mu_0}{\gamma}\right)^3 \rm{ln}\left(1 + 
\frac{\gamma}{\mu_0}\right) - \frac{3}{4} \frac{\mu_0}{\gamma} 
e^{-\gamma\tau/\mu_0} \right].
\label{irr_grey}
\end{equation}
Here $\mu_{0}$ is the cosine of the angle between each point on the planet's 
surface and the illuminating star; $T_{eq}$ is the planet's equilibrium
temperature; $T_{eff}$ describes the additional contribution due to internal
heating such that $T^4 = T_{eff}^4 + T_{eq}^4$ at optical 
depth 2/3 and $\mu_{0} = 1$; and $\gamma$ is the ratio of the 
optical depth of absorption to that of emitted radiation, which parameterizes
the deposition of stellar energy into the planetary atmosphere.  A 
self-consistent value for the factor
$\gamma$ is determined iteratively after computing the emission and 
transmission spectra of the planet (Section~\ref{spectra}).  Our choices
for the other model inputs to Eq.~\ref{irr_grey} are as follows.  Since we do 
not model the scattering 
properties of the planetary atmosphere, we must assume
values for the planet's albedo and its implied $T_{eq}$.  We employ an  
albedo of 0.15 for cloud-free
models and 0.65 for models with sulfuric acid clouds, resulting in  
values of $T_{eq}$ = 368 K and 295 K respectively.  Despite the
presence of additional planets in the Gl 581 system, tidal heating due to 
these bodies should not add significantly to the planet's observed
temperature (Wade Henning, private communication), and other sources
internal heating should be small.  However, our temperature model requires
that a small but non-zero value be employed for the internal heating via
the parameter $T_{eff}$, to ensure a 
physically reaslistic temperature profile at high optical depths.  We choose
a value of $T_{eff} = 50$ K, which adds negligibly to the global energy emitted
by the planet at optical depth of unity.

The pressure structure of our super-Earth atmosphere is calculated by 
integration of the equation of hydrostatic equilibrium
\begin{equation}
\frac{\rm{d}P}{\rm{d}\tau} = \frac{g}{\kappa_{gr}},
\end{equation}
where $\kappa_{gr}$ is the Planck mean opacity in units of cm$^2$/g.

\begin{figure}
\plotone{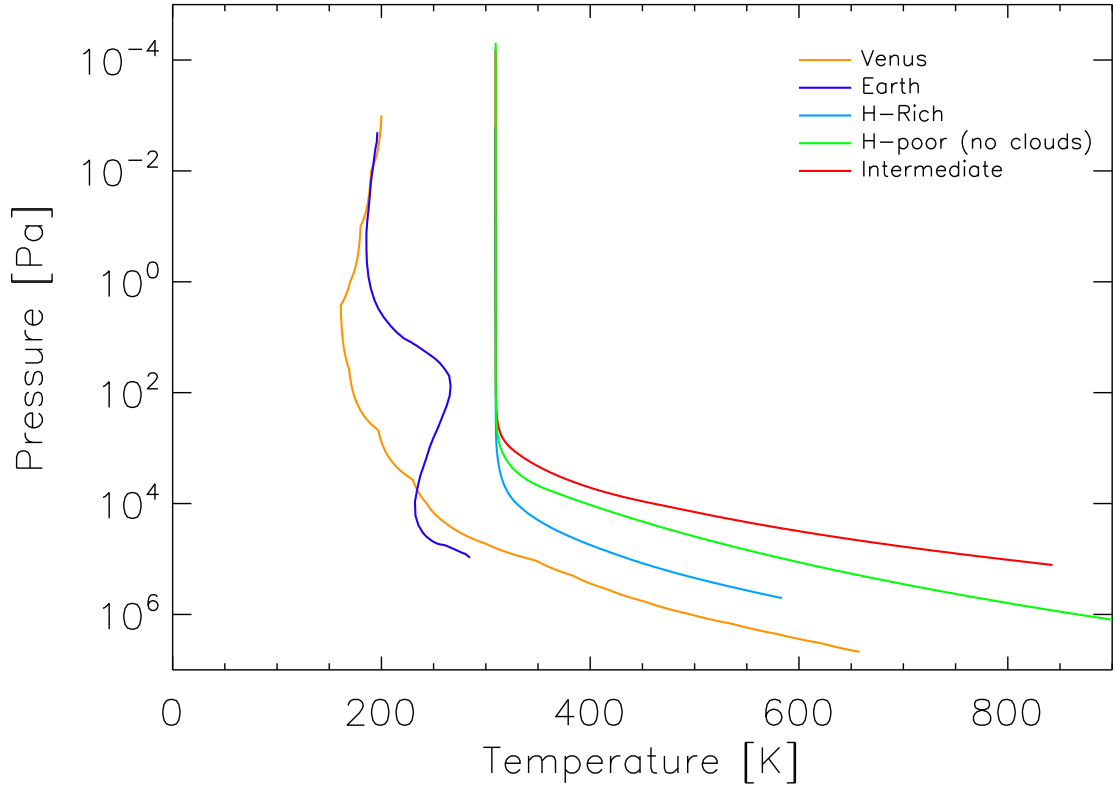}
\caption{Temperature-pressure profiles for each of our three atmosphere 
        scenarios (at $\mu_0 = 0.5$) along with those of Earth and Venus as a 
	reference.  
	Qualitatively, our T-P profiles resemble Venus', however our model
	is not capable of reproducing the types of temperature inversions 
	seen in the upper atmospheres of the terrestrial planets.
        \label{t_p_fig}}
\end{figure}

We check the resulting radiative temperature-pressure profile for
instability to convection according to the stability criterion
\begin{equation}
\frac{\rm{d}\rm{ln}T}{\rm{d}\rm{ln}P} < \nabla_{ad},
\end{equation}
where $\nabla_{ad} = (\gamma - 1)/\gamma$, and $\gamma$ is the adiabatic 
index.  If this criterion is satisfied then we employ the radiative T-P 
profile described above.  
Otherwise, we assume that the atmospheric T-P profile follows an adiabat with
the appropriate lapse rate.  The resulting temperature-pressure profiles
for the atmospheres described in the following section are shown in 
Figure~\ref{t_p_fig} along with those of Venus and Earth as a reference.

\subsection{Atmospheric Mass}

The mass of the Gl 581c's atmosphere is one of the major unknowns 
involved in any model.  In our own Solar System, the atmospheric
masses of similarly-sized Earth and Venus vary by almost two orders of
magnitude, which implies that other factors such as formation history 
or the presence of liquid oceans have a large
effect.    Fortunately, the results we present in this paper have little 
dependence on the mass of the atmosphere (see Section~\ref{results}).  

For our models, we pick an atmospheric mass and then ensure 
consistency with our chosen 
optical depth scale.  We determine the atmospheric density, $\rho(r)$, as a 
function of height, given our T-P profile and the chemistry that is implied, 
using an ideal gas equation of state.  
The total mass contained in the atmosphere, $M_{atm}$, is then calculated by 
integrating 
the equation of conservation of mass upward from the planet's surface through 
the atmosphere
\begin{equation}
M_{atm} = \int 4 \pi r^{2} \rho(r) \,dr,
\end{equation}
where $r$ is the distance from the planet's center.  Given the atmospheric 
mass that we have decided upon for a given model, we iteratively 
adjust the value for $\tau$ at the base of the atmosphere until our calculated
atmospheric mass agrees with the value of the model input.  In 
Section~\ref{comp} we discuss our choices of mass for the different
atmosphere scenarios.  

\subsection{Emission and Transmission Spectra \label{spectra}}

To determine the planet's emitted spectrum, we integrate the equation of
radiative transfer through the planet's atmosphere.  Using the T-P profile
as outlined in Section~\ref{t_p}, we obtain the emergent 
intensity, $I$, by
\begin{equation}
I(\lambda,\mu) = \frac{1}{\mu}\int^{\tau}_{0} S(T) e^{-\tau'/\mu} \,d\tau',
\end{equation}
where S is the source function, $\mu$ is the cosine of the viewing angle, and
$\tau$ is the optical depth at the base of the atmosphere.  
We have made the simplifying assumption of a Planckian source function.  We 
also do not consider scattering, which is a valid assumption in the IR where 
this is generally a small effect.  We have tested the validity of both of
these assumptions by using this scheme to reproduce Earth and Venus' emitted 
spectra, given
their known T-P profiles.  We simulate Venus' opaque H$_2$SO$_4$ cloud deck by 
cutting off the planetary emission at an altitude of 70 km above the surface.
Our resulting spectra are in agreement with the expected planet-averaged 
emission for both of these cases \citep{mor86, han72}. 

If the planet's orbit is aligned such that it transits its host star, then the 
atmospheric transmission spectrum 
can potentially be observed.  During transit, light from the star passes 
through the
optically thin upper layers of the planet's atmosphere, leading to excess
absorption of star light at certain wavelengths.  We compute theoretical 
transmission spectra by determining the amount of 
absorption from light rays passing through the planet's 
atmosphere, along the observer's line of sight.  The geometry we employ is 
similar to the one presented in \citet[][their Figure 1]{ehr06}, where the 
solid core of the planet is represented by an optically thick disk of radius 
1.4 R$_{\earth}$.  We calculate transmission through the annulus of gas
surrounding this disk, which represents the planet's atmosphere.  The 
resultant intensities are then given by the expression
\begin{equation}
I(\lambda, \mu) = I_{0} e^{-\tau/\mu},
\end{equation}
where $I_0$ is the incident intensity from the star and $\tau$ is now the 
slant optical depth integrated through the planet's atmosphere along the 
observer's line of 
sight.  We once again only include absorption here and do not include the
effects of additional scattering out of the beam, which we find to have
a negligible effect longward of 1 $\mu$m.  We then integrate over the 
entire annulus of the atmosphere to determine the total excess 
absorption of stellar flux as a function of wavelength.  

To calculate transmission 
spectra, it is necessary to assume a planetary radius,
R$_p$.  This is required both to determine the amount of stellar flux that 
is blocked out during transit
as well as for computing path lengths through the planet's atmosphere.  Since
there are no transit observations available for Gl 581c,
R$_p$ becomes a free parameter in our model.  We therefore rely on theoretical
predictions of the mass-radius relationship for super-Earths 
and choose a radius of 1.4 
R$_{\earth}$, corresponding to a rocky planet with a composition similar to 
Earth, composed of 67.5\% silicate mantle and 32.5\% iron core \citep{sea07, 
val07}.  

\section{Atmospheric Composition \label{comp}}

The formation history of Gl 581c is largely unknown, and similarly, the
mechanism by which the planet obtained its atmosphere remains uncertain.  
In particular, for the terrestrial planets in our Solar System, the issue of 
atmospheric
escape plays a central role in our understanding of these planets' histories.
Since estimates of atmospheric escape rates for hot giant exoplanets 
are uncertain to several orders
of magnitude \citep{bar04, hub07}, it is not clear 
whether Gl 581c and similarly sized Super-Earths would be able to retain 
large amounts of molecular hydrogen in their atmospheres.  

In the thermal limit, the ability of an atmosphere to retain hydrogen is
described by the escape parameter
\begin{equation}
\lambda = \frac{R_{pl}}{H}, \label{lambda}
\end{equation} 
where $H$ is the atmospheric scale height for hydrogen at the exobase.  The 
factor $\lambda$ is also 
equivalent to the square of the ratio between the escape speed and mean thermal
speed of the gas.  High values of $\lambda$ ($\gtrsim 15$) mean that the 
atmosphere
will be retained, while planets with low values will experience atmospheric 
escape.  For reference, in Earth's atmosphere $\lambda$ is between 5 and 10
for hydrogen.  For Gl 581c, the planet's high escape speed potentially
implies an escape parameter larger than $15$.  However,
our atmosphere model does not include the UV information necessary to
predict exobase temperatures, and at present the XUV flux from Gl 581 is 
unconstrained.  

From this argument, thermal escape of atomic hydrogen of Gl 581c
may be unlikely.  However, a variety of nonthermal escape processes can 
procede at much faster rates than thermal escape.  These are known to 
dominate in many cases \citep{hun89} and are more difficult to 
constrain.  Gl 581c 
therefore lies in a region of parameter
space where it is unclear whether the planet is able to retain hydrogen over
its lifetime.  As a further complication, the initial amount of outgassing of 
hydrogen and other gasses from the planetary surface is
almost entirely unconstrained, making it difficult to ascertain how
hydrogen loss would contribute to the present atmospheric composition 
\citep{elk08}.  

Since the amount of atmospheric hydrogen is therefore poorly constrained, and 
in order to present results of more general relevance,
we choose to model three distinctly different atmospheres for Gl 581c.
These vary in their composition and in the ways by which they would
be acquired by the planet.  The three cases that we choose also span the
range from hydrogen-rich to hydrogen-poor in our effort to understand
the different observational signatures of these types of atmospheres.  
The scenarios are outlined as follows.

\subsection{Hydrogen-Rich Atmosphere} 

Our first model atmosphere represents one obtained
through a standard accretion process that experiences little evolution later 
in its lifetime.  In this scenario, after formation of 
a solid core, Gl 581c would accrete its atmosphere through gravitational 
capture of gas from the protoplanetary disk, resulting in a composition
that more or less reflects that of the stellar nebula.  It is assumed that
this atmosphere is sufficiently massive and experiences low escape rates, 
such that the planet is able to retain large quantities of hydrogen.  This 
results in a reducing chemistry where hydrogen-bearing molecules dominate 
the atmosphere.

\begin{figure}
\begin{center}
\includegraphics[scale=0.64]{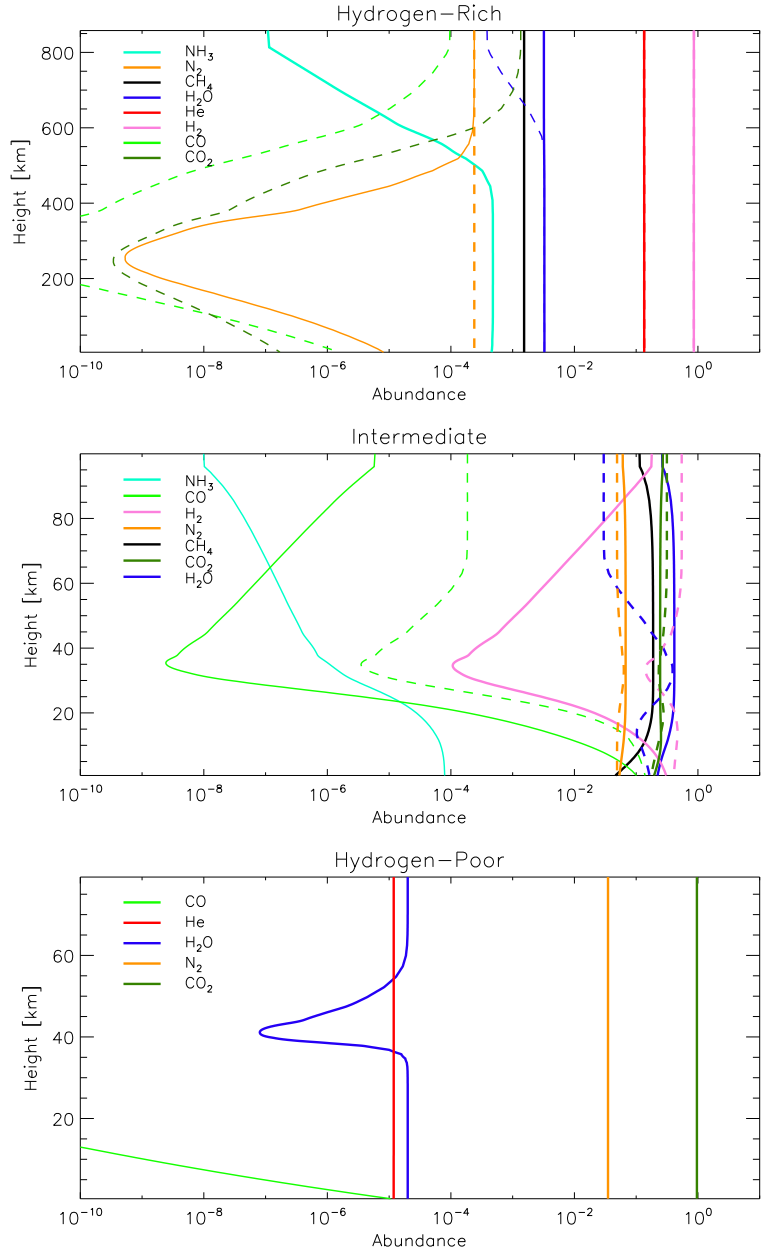}
\end{center}
\caption{Molecular abundances as a funcion of height for each of the
        three super-Earth atmosphere scenarios.  The solid lines show the
	results of the chemical equilibrium calculations, while the dashed
	lines give molecular abundances for the case that NH$_3$ and CH$_4$
	have been entirely removed from the atmosphere through 
	photodissociation.  In the top panel, the solid and dashed lines for
	both H$_2$ and He lie essentially on top of one another, since their
	abundances barely change with the loss of ammonia and methane from the
	atmosphere.
        \label{chem}}
\end{figure}

To calculate the detailed composition for the atmosphere of Gl 581c we 
use the observed architecture of the Gl 581 planet system to infer that
all three Gl 581 planets formed at about 2-5 times larger orbital radii
and migrated to their current locations. This would imply that both Gl 581c 
and d
had access to volatiles during formation beyond the snow line, similarly to
the HD 69830 system with three Neptunes described in \citet{lov06}.
For HD 69830 \citet{bei05} have done IR spectroscopy of the debris disk between
the orbits of HD 69830c and d, finding a volatile mix corresponding to a disk
producing cometary body like Hale-Bopp, with its known bulk composition
\citep{irv00}. By analogy, we assume the same bulk composition for Gl 581c, or
any super-Earth planet under similar conditions. 
This calculation results in an atmosphere of Gl 581c that is
composed by mass
of 73.2\% H, 22.6\% He, 3.5\% C, N, and O, 0.2\% Ne, and 0.5\% other heavy 
elements appearing in solar composition ratios (see Table~\ref{composition}).  
To determine the 
resulting molecular composition of the atmosphere we apply the conditions of 
chemical equilibrium as discussed in Section~\ref{chemistry}.  The resulting 
mixing ratios for the major constituents of this atmosphere are shown in the 
top panel of Figure~\ref{chem} (solid lines).  

Due to its large hydrogen content, the chemical equlilbrium calculation 
requires that this atmosphere have high abundances of hydrogen-bearing 
molecules such as H$_2$,
H$_2$O, CH$_4$, and NH$_3$.  However, as mentioned in Section~\ref{chemistry},
ammonia and methane are apt to be destroyed by UV radiation from the star.  
We determine the 
photochemical lifetimes of NH$_3$ and CH$_4$ as a function of height above
the planet's surface according to  Eq.~\ref{photochem}, and we determine that 
UV photodissociation 
should readily remove ammonia and methane from the atmosphere as long
as any chemical reactions that create these molecules occur slowly.  We 
therefore include a photochemical model for the hydrogen-rich scenario,
where NH$_3$ and CH$_4$ are completely removed from the atmosphere along with
the chemical equilibrium model, as
shown in Figure~\ref{chem}.  

We employ an atmospheric mass for this model that is scaled up by a factor of 
4 compared to that of
the Earth, to account for the large quantity of hydrogen that remains in this
atmosphere.  However, this number remains somewhat arbitrary since the true 
mass of the atmosphere for this scenario is dependent on
the amount of gas that can be gravitationally captured and then retained by a 
planet of this size.  Our choice of atmospheric mass has indirect implications 
for 
photochemistry as well in that a more massive atmosphere may have some 
remaining CH$_4$ and NH$_3$ present if these species are 
only readily destroyed in the upper atmosphere.

\subsection {Intermediate Hydrogen Content Atmosphere} 

Our second case falls midway between a hydrogen-rich atmosphere and
one that has lost all of its hydrogen through escape processes.  We work under 
the assumption here that Gl 581c has experienced atmospheric evolution to 
the point that it has lost a large amount of the hydrogen obtained in the
accretion process.  However, in this scenario, the atmosphere
has still managed to retain some hydrogen so that the atmospheric chemistry 
remains mildly reducing.  This type of atmospheric chemistry does not occur
in any Solar System bodies, however it is possible that such an atmosphere
could take hold on Gl 581c via a number of methods.  One
possibility is that atmospheric escape occured early on in the planet's life
when Gl 581 was still in a young active phase, but that the escape
process was cut off prematurely as the M-star evolved and produced a reduced 
UV and X-ray flux.  This is similar to what is known to occur in solar-type
stars as they age, where UV flux reduces by a factor of 10$^3$ over their 
main-sequence lifetimes \citep{rib05}.  A second
possibility is that the planet did lose its entire atmosphere
early on in its history through hydrodynamic escape, and that this represents 
a secondary atmosphere
obtained through planetary outgassing.  Large quantities of H$_2$ can 
potentially be outgassed through the process 
H$_2$O + Fe $\rightarrow$ FeO + H$_2$ \citep{elk08}, and much of this 
hydrogen may be retained due to the high surface gravity of the super-Earth.   

We choose to work under the first assumption that Gl 581c has experienced
incomplete loss of its atmospheric hydrogen.  The outgassing scenario is viable
as well, although it is difficult to predict what the resulting atmospheric
composition would resemble without a more complete picture of the interior
composition and formation history of the planet.  We therefore assume that 
Gl 581c is not an ocean planet and that the lightest constituents of the 
atmosphere have been able to escape during the lifetime of the planet.
Our evaporated case then develops in the direction of
Mars' atmosphere with He and Ne becoming trace constituents.
The atmospheric hydrogen would mostly escape as well, and what 
remains is locked in heavier molecular species (mostly H$_2$O and CH$_4$).  

To determine the atmospheric composition for this scenario, we start with the 
hydrogen-rich atmosphere described 
above and then assume that a portion of the hydrogen is allowed to escape
to the point that it is reduced to half of its initial abundance.  
With what remains we then recalculate chemical equilibrium to determine the 
resultant composition of the atmosphere.  
This process results in an 
atmosphere that has experienced a partial loss of hydrogen, but still leaves 
behind a reduced but non-negligible amount of molecular hydrogen in the 
planet's atmosphere (about 1-10\% by volume).  The rest of the atmosphere is 
made up mostly of water 
vapor, CO$_2$, CH$_4$, and N$_2$.  Photochemistry once again affects the upper 
atmosphere in this scenario in that methane and ammonia are 
photochemically unstable.  We therefore remove these two molecules from
our model, and we present this case along with the case of chemical
equilibrium in Figure~\ref{chem} (middle panel).  For this scenario we have
chosen an atmospheric mass equivalent to that of Earth's atmosphere.

\subsection {Hydrogen-Poor Atmosphere}

For our final scenario, we examine what the atmosphere of Gl 581c would 
resemble if it had a similar formation history to those of the terrestrial 
planets in our Solar System.  We choose a composition mirroring that of Venus' 
atmosphere, where it is assumed that there has been an almost complete loss of 
hydrogen through molecular photodissociation and subsequent escape.  
For Venus, the planetary surface is too warm for liquid 
H$_2$O, and carbon remains free in the atmosphere rather than being sequestered
to the bottom of a water ocean.  This is most likely the case for Gl 581c as 
well, since the planet is expected to have a high surface temperature 
\citep{sel07}.  The 
oxidizing nature of this atmospheric chemistry then leads to the formation of 
large quantities of CO$_2$, which becomes the primary source of opacity.  
To keep with the comparison to Venus, we have also given this atmosphere a mass
equal to Venus' atmospheric mass, or about 100 times greater than that of the
Earth.  We assume that this planetary atmosphere has the same atomic makeup as 
Venus (Table~\ref{composition}), and the resulting equilibrium molecular 
composition as a function of height is shown in Figure~\ref{chem}.

For a Venusian atmosphere, photochemistry plays a role in 
aiding in the formation of sulfuric acid clouds via the reaction
2SO$_2$ + O$_2$ + 2H$_2$O $\rightarrow$ 2H$_2$SO$_4$.  As long as the planet's 
atmosphere contains some SO$_2$ this process will occur, and under the right 
conditions this will result in a cloud layer on Gl 581c just like the one on 
Venus.  Super-Earths, being more massive than Venus, are very likely to outgas 
larger amounts of SO$_2$ and H$_2$S than Venus did. With loss of H in this 
scenario, the concentration of SO$_2$ will be larger than what it is in Venus. 
About 50\% of the sulfur will be outgased as H$_2$S \citep{wal94}.
For sulfuric acid clouds to exist on Gl 581c, the partial pressure of
H$_2$SO$_4$ must be sufficiently large such that its temperature-pressure
curve in the atmosphere crosses the condensation 
curve.  For this scenario, condensation will then occur if the abundance of 
sulfuric acid excedes 0.5 parts per
million (ppm).  For reference, on Venus the H$_2$SO$_4$ abundance just
below the cloud deck is approximately 10 ppm.  For Gl 581c, 
sulfuric acid abundances ranging from 3 ppm to 0.3\% will result in clouds
ranging in location from P = 0.15 bar (T = 400 K) to 25 km P = 25 mbar 
(T = 300 K) above the planet's surface.  Due to the uncertainties in this
cloud model, in the following sections we present
models both with and without sulfuric acid clouds for the hydrogen-poor 
scenario.

\section{Results \label{results}}

\subsection{Transmission Spectra as a Probe of Hydrogen Content}
Our most significant finding is that transmission spectra are the best
method for distinguishing between hydrogen-dominated atmospheres and 
hydrogen-poor
atmospheres. For a conceptual explanation, we can consider an exoplanet
atmosphere to have a vertical extent of 10 $H$, where $H$ is the scale height,
\begin{equation}
H = \frac{kT}{\mu_m g}.\label{H_eq}
\end{equation}
The scale height,  
is proportional to the inverse of the atmospheric mean molecular weight 
$\mu_m$, expressed from here on in terms of atomic mass units (u).
In hydrogen-dominated atmospheres, $\mu_m \simeq$ 2, and $H$ is
large.  In contrast, in our hydrogen-poor atmosphere, $\mu_m \simeq 40$, 
making the scale height, and 
hence the atmosphere available for transmission, less by a factor of 20.  
Measuring the change in eclipse depth $\Delta D$ across spectral lines in 
transmission gives an order of magnitude approximation of the scale height 
xaccording to
\begin{equation}
\Delta D \sim \frac{2 H R_{pl}}{R_{*}^{2}}, \label{depth}
\end{equation}
allowing for a determination of 
$\mu_m$ assuming that both the atmospheric temperature and surface
gravity are known to within a factor of a few.  This will indeed be the
case as long as we assume that the 
equilibrium temperature of the planet predicts its 
actual temperature to within a factor of several and that the radius and mass
of the planet are known, allowing for a calculation of $g$.  Measurement of 
the scale 
height should therefore offer a strong constraint on the atmospheric 
composition in terms of its hydrogen content, via the parameter
$\mu_m$, even if the exact T-P profile of the atmosphere is unknown.  This 
interplay between scale height and transmission signal has also been pointed 
out by \citet{ehr06} for low-mass transiting planets.

\begin{figure}
\plotone{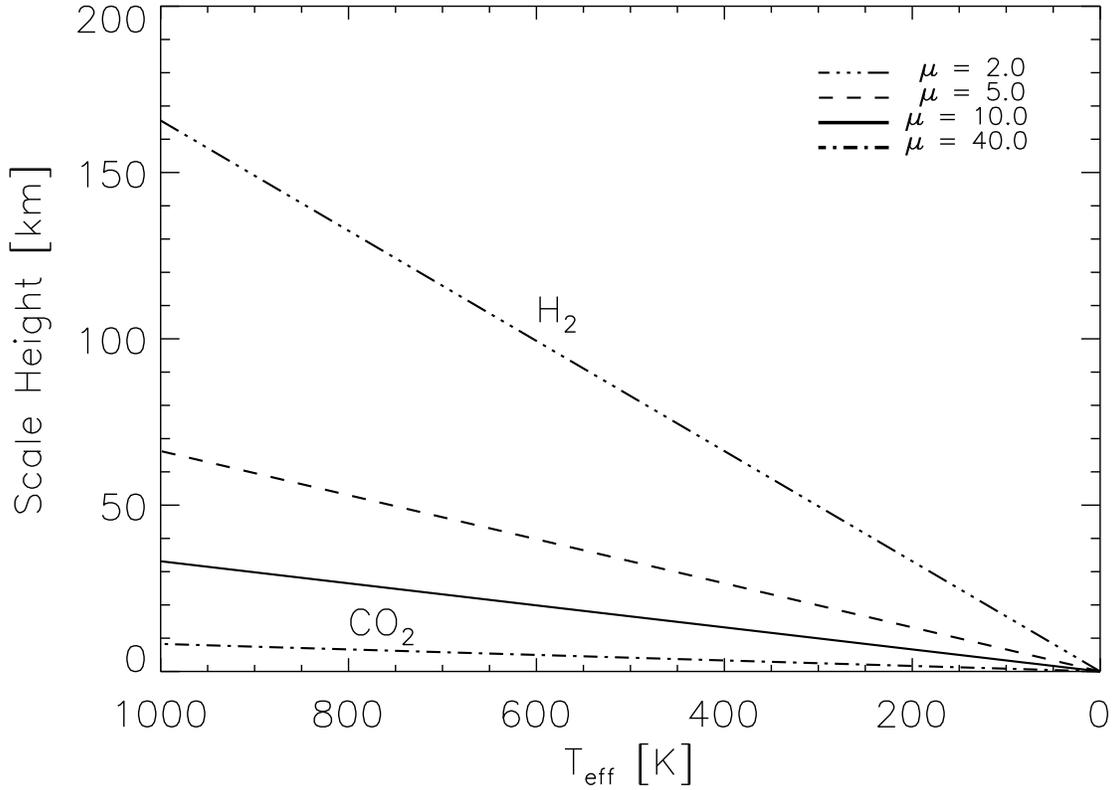}
\caption{Atmospheric scale height as a function of effective temperature
        for a number of atmospheric compositions.  Each line corresponds to a
	different value of mean molecular weight, $\mu_m$.  At the top,  
	$\mu_m = 2$ signifies an atmosphere comprised solely of molecular
	hydrogen, while the bottom line with $\mu_m = 40$ represents an 
	atmosphere made up of mostly CO$_2$.  The observability of an
	exoplanet atmosphere in transmission falls off as molecular weight 
	increases and scale height correspondingly drops off.  This plot is
	for Gl 581c, assuming a surface gravity of 25 m/s$^2$.
        \label{scale_height}}
\end{figure}

The sensitivity of scale height to
atmospheric composition  is illustrated in Figure~\ref{scale_height}, where 
we show scale height as a function of effective temperature for a variety of 
different $\mu_m$.  The implication is that, for the same mass of atmosphere, 
one that is made predominantly of hydrogen will tend to be larger and 
therefore more readily observable in transmission.  For the other two cases 
where the atmosphere is composed predominantly of heavier molecules, we 
have the opposite case.  Additionally, since scale height drops
off as $T_{eff}$ decreases, planets that are farther away from their host
stars (and hence colder) will also be difficult to observe in transmission.  
This has implications
for follow-up of transiting giant planets discovered by the Kepler 
\citep{bor04, bas05} mission, since they will detect planets with orbital 
periods of a year or more. 

In Figure~\ref{transmission} we have plotted our modeled spectra for each
of the three cases of atmospheric composition that were discussed in 
Section~\ref{comp}.
As expected from the scale height argument presented above, the hydrogen-rich
atmosphere shows deep absorption features in transmission at a level of up to
$10^{-4}$.  The other two atmospheres have weaker
features -- by about a factor of 10 for the intermediate atmosphere, and even 
weaker for the hydrogen-poor atmosphere.   In the latter cases, due to the 
small scale height, the transmission spectrum only probes a very narrow range 
of height in the atmosphere.  This result holds true independently of the T-P 
profile that we employ.

Ideally, with the capability to detect transmission spectra at a level of
less than a part in $10^{-5}$ relative to the star, we could learn about the 
hydrogen content of a super-Earth atmosphere like Gl 581c.  IR transmission 
spectroscopy 
would be the easiest way to differentiate between the various atmospheric 
cases, however narrow-band photometry from Spitzer is currently the best that 
can be accomplished.  In Figure~\ref{transmission} we also show the expected
relative fluxes in each of the Spitzer IRAC bands and in the MIPS 24-$\mu$m
band for each model atmosphere.  Each of the Spitzer bands averages over
a number of spectral features, which effectively smooths out much of the 
transmission signal.  However there are perceptible changes
from one channel to the next that could be suggestive of atmospheric 
composition and hydrogen content.  In the case of the hydrogen-rich atmosphere,
the expected Spitzer IRAC fluxes vary from band to band at a level of 
10$^{-5}$, while
in the intermediate and hydrogen-poor cases this is reduced to a factor of 
10$^{-6}$.  For other exoplanet systems the expected transmission signal
will vary according to Eq.~\ref{depth}.  If photometry at this level can be 
obtained, which will most likely need to
wait for the launch of JWST, it will give observers a chance
to place meaningful constraints on the atmospheric composition of super-Earths
like Gl 581c.  

Changing the atmospheric mass does little to alter the result presented 
above.  In general, increasing the mass of the atmosphere serves to 
raise the height above the planet's surface at which the atmosphere
becomes optically thick.  Since one can only see through optically thin layers,
the result will be an overall transit radius that 
appears larger.  This can be seen in Figure~\ref{transmission}, where the 
hydrogen-rich atmosphere is the most massive one that we considered, and its
transit radius appears somewhat larger than for the other cases by a 
factor of a few parts in 10$^{-5}$.  However, the scale height itself,
measured from the relative depth of the spectral lines, has very 
little dependence on atmospheric mass unless either chemistry or 
temperature changes dramatically as a function of height through 
the atmosphere.   In summary, the depth of any spectral features observed in 
transmission should remain mostly independent of the atmosphere's mass, and 
should depend most strongly on the scale height, $H$.  This is an encouraging 
result, since the total
mass of an atmosphere cannot be known a priori, and the values that we have
assigned for atmospheric mass in this paper have been somewhat arbitrarily
scaled from terrestrial planets in our Solar System.  A negative corollary 
however, is that the 
mass of the atmosphere generally cannot be determined from the planet's
transmission spectrum.

\subsection{Thermal Emission Spectrum}

The planetary emission spectrum can also be generally helpful in discerning 
the chemistry and composition of a Super-Earth atmosphere.  We find
that the
three types of atmosphere that we explored for Gl 581c each reveal very 
different emission spectra.  The observed features in a planet's emission 
spectrum represent the dominant sources of opacity in that 
atmosphere - H$_2$O for our hydrogen-dominated atmosphere, CO$_2$ for the
hydrogen-poor atmosphere, and a combination of both in the intermediate 
hydrogen content case.   The spectral features are therefore a strong 
indicator of
the chemistry at work in the planetary atmosphere.  Figure~\ref{emission} 
shows the wavelength-dependent emission 
calculated for each of the three 
cases of atmospheric composition that we have explored.  We have indicated 
the effect of methane and ammonia photochemistry on the emission spectra for 
the hydrogen-rich and
intermediate cases, and the effect of a possible sulfuric acid cloud layer in 
the hydrogen-poor case.  

In the hydrogen-rich case, atmospheric absorption lines are
almost exclusively due to water.  While methane and ammonia are not expected to
be photochemically stable in this atmosphere, their outstanding presence would 
affect the
emission spectrum between 5 and 15 $\mu$m.  Supression of flux in the 7 to 9
$\mu$m range is indicative of methane, and a strong feature between
10 and 11 microns could indicate the presence of ammonia.  

The intermediate atmosphere with its mildly reducing chemistry exhibits 
strong features due to both water and CO$_2$ absorption.  The effect of ammonia
photochemistry on this spectrum is minimal, while the presence or absence
of methane can have a much larger impact.  A complete loss of methane from
this atmosphere results in a small spectral window from 2 to 3 $\mu$m where 
there is little to no opacity, and the atmosphere remains optically thin down
to the planet's surface.   However, flux in this range can be further supressed
due to the presence of aerosols in the atmosphere,
or by a lower surface temperature than the one employed in our model.

For the hydrogen-poor atmosphere the presence of clouds has a strong effect
on the emission spectrum between 1 and 9 $\mu$m.  In this scenario
CO$_2$ is the predominant source of opacity, resulting in absorption features
at 4.5 $\mu$m, 11 $\mu$m, and 15$\mu$m.  Weak water features are also
present at longer wavelengths despite the low abundance of H$_2$O in this
atmosphere.  In the absence of clouds, there are few absorption features in 
the 1-8 $\mu$m range, and the atmosphere remains optically thin almost down to
the planet's hot surface.   However, the presence
of opaque clouds can greatly supress the flux in this spectral range.  
Additionally, other sources of opacity between 1 and 
8 $\mu$m  that have not been accounted for in our model such as
aerosols or photochemical hazes could supress the 
emitted flux in this spectral range.  For the cloudy hydrogen-poor model, 
we have placed an opaque cloud deck at a temperature of 400 K (P = 200 mbar).  
Moving this to 
higher in the atmosphere would result in an even lower emitted flux in 
the near-IR due to the lower temperatures in this region of the atmosphere.

To date, emission from extrasolar planets has been best measured
through observations of secondary eclipses for transiting systems (when the
planet passes behind its host star).  The depth $D$ of secondary eclipse can 
be expressed as  
\begin{equation}
D = \frac{F_{pl}}{F_{*}+F_{pl}}
\end{equation}
where $F_{pl}$ and $F_{*}$ are the planetary and stellar fluxes respectively.
In Figure~\ref{color_mag} we plot the secondary eclipse depths for our
three model atmospheres in the 1-30
$\mu$m range.  For the stellar spectrum we use a Kurucz 
model\footnote{See http://kurucz.harvard.edu/grids.html} with stellar 
parameters $T_{eff}$ = 3500 K, 
log $g$ = 5.0, and [Fe/H] = -0.3.
The three very differently shaped emission spectra for our atmosphere
scenarios also manifest themselves in differing secondary eclipse depths, where
the same spectral features from the planetary emission spectra in 
Figure~\ref{emission} can be seen in Figure~\ref{color_mag} as well.  The 
expected eclipse depths for Gl 581c in the IR are all predicted to be
less than 70 ppm with the exception of the cloud-free hydrogen-poor
model.

In Figure~\ref{color_mag} we also show the secondary eclipse depths 
for our three model atmospheres in each of the Spiter IRAC 
bands and in the MIPS 24 $\mu$m band.  The IRAC bands give good coverage of 
water features in both the hydrogen-rich and intermediate atmospheres,
although the expected eclipse depths are quite small at less than 10 ppm.  For
the cloud-free hydrogen-poor atmosphere we predict large variations in 
eclipse depth between the IRAC channels, indicating the presence of CO$_2$.  
However, for the cloudy hydrogen-poor model, features in this spectral
range are washed out, and the IRAC bands simply sample the blackbody curve at 
the temperature of the top of the cloud deck.  In terms of
constraining the nature of the planet's atmosphere, the detection of water 
features in the 1 - 10 $\mu$m range and the CO$_2$ feature in the 12-19 
$\mu$m range would be highly indicative of the atmospheric chemistry at
work.  Additionally, detection of other hydrogen-bearing molecules
such as CH$_4$ and NH$_3$ would be of interest as they point to the type
of photochemical processes taking place in the planet's atmosphere.  Already
methane has been detected in the atmosphere of a giant exoplanet at levels
that are not easily explained with current models \citep{swa08}. 
Unfortunately, with the exception of the cloud-free hydrogen-poor model, 
detecting spectral features from secondary eclipse observations would 
necessitate observations with a precision
of better than 10$^{-5}$, which is not yet feasible with current instruments.
This factor becomes larger or smaller depending on the stellar and planetary
radii and temperatures, since warmer planets and larger planets such as ocean
planets are easier to detect.  
Additionally, Spitzer does not carry any instruments that 
are capable of broad band photometry between 10 and 20 $\mu$m, which 
effectively means that the telescope misses the large CO$_2$ feature at 15 
$\mu$m. 
The types of measurments suggested by Figure~\ref{color_mag} will therefore 
most likely need to await future instruments such as MIRI \citep{wri03} 
aboard JWST. 

Temperature inversions, where the atmospheric temperature
\textit{increases} as a function of height, have the potential to greatly alter
the emission spectrum.  However, the locations of the various spectral
features (in terms of wavelength) are robust regardless of the
temperature gradient and will therefore remain good indicators of atmospheric
chemistry.  In general, a temperature inversion at the right
height in the atmosphere will result in a spectrum of emission lines
rather than the absorption features presented in Figures~\ref{color_mag} 
and~\ref{emission}.  Spectra for atmospheres with inverted temperature
profiles will therefore look quite different from the ones presented here,
and for this reason emission spectra are also a strong indicator
of the atmospheric temperature gradient.

Atmospheric mass can have a stronger effect on emission spectra than on 
transmission, although once again the effect is secondary.  A larger
atmospheric mass will result in a higher surface temperature, since the 
density of gas at the base of the atmosphere is greater.
In the optically thin regime, a higher surface temperature translates to 
a larger overall continuum flux, so a more massive atmosphere could result in
an emission signature that is easier to observe.  Additionally, 
increasing the
mass of the planet's atmosphere can cause a shift from the optically thin
to optically thick regime, which in turn can significantly alter the 
appearance of its emitted spectrum.  This would apply more strongly to the 
visible
portion of the spectrum however, since the atmosphere tends to be mostly
optically thick in the IR.  Additionally, if the changes in surface temperature
and pressure between a massive atmosphere and a low-mass one are large
enough to alter the atmospheric chemistry, then the absorption features in the 
emission spectrum will have a mass dependence as well.  In terms of 
photochemistry, a more massive atmosphere has a higher likelihood of 
containing stable quantities of methane and possibly ammonia, which affects 
the emission spectrum as seen in Figure~\ref{emission}.

\section{Discussion \label{discussion}}

We have determined a method for differentiating between Super-Earths
with large hydrogen-rich atmospheres and those with atmospheres composed
mostly of heavier molecules.  Since the relative depth of spectral features in 
a planet's transmission spectrum results in a direct determination of the 
atmospheric scale height, 
the mean molecular weight of that atmosphere can then be calculated by a 
simple relation (see Eq.~\ref{H_eq}).  For the case of Gl 581c, a 
hydrogen-rich atmosphere with its large scale height reveals spectral lines in 
transmission at a level of 10$^{-4}$.  On the other hand, the transmission 
features for a hydrogen-poor atmosphere will be more than an order of
magnitude smaller at a level of less than 10$^{-5}$.  The presence
of clouds will not alter this conclusion unless they are
unrealistically high in the atmosphere - at a height
approaching 10 scale heights above the planet's surface, where the 
atmospheric density has dropped off to a point that is is difficult to 
imagine the presence of an optically thick cloud deck.
We therefore emphasize
that transmission spectroscopy should be a robust method for determining
the atmospheric scale height.

The emission spectrum should also be generally helpful in constraining
atmospheric hydrogen content.  However interpretation can be more
complicated than for transmission spectra due to the dependence of the 
emission spectrum on the atmospheric temperature gradient.  For a 
hydrogen-dominated super-Earth, the emission
spectrum should be marked mostly by water features.
As we then transition in the cases that we have presented, from H-rich to 
H-poor, we find that the planetary emission starts showing weaker H$_2$O 
features, while
absorption due to CO$_2$ increases.  Additionally, the presence of methane or
ammonia features for hydrogen-containing atmospheres could be an interesting
indication of the photochemistry at work, since these molecules are thought to
be destroyed by photodissociation over the lifetime of the planet.  
Even a weak detection of these features could therefore be helpful in 
constraining models for atmospheric evolution.

The exact mass of Gl 581c's atmosphere, which has been chosen somewhat 
arbitrarily in this paper, has little effect on our spectra presented
above.  This is due to the simple fact that regions beyond an optical depth
of unity are essentially hidden from observers.  Therefore, as long as 
the atmosphere becomes optically thick at some height above the planetary
suface, a more massive atmosphere 
would  reveal the same emission and transmission spectra as a less massive
one barring any secondary effects such as differing cloud structures.  
Pressure and temperature conditions near the planetary surface, 
which would be indicative of the
atmospheric mass cannot be observed via spectroscopy for an optically thick
atmosphere.  For this reason it will be difficult to constrain atmospheric
mass with spectral observations.  Another negative consequence is that
the question of habitability will remain unanswered as long as the 
planet's surface
conditions are unknown and unobservable.  Mass-radius measurements may
provide the first constraints on atmospheric mass for super-Earths, however 
there are large degeneracies in the models that will be difficult to overcome 
\citep{ada07}.

Our simple model for the super-Earth atmosphere has allowed us to 
understand some of the important effects of hydrogen content on transmission
and emission spectra.  However, future more detailed models will be needed to 
address some of the issues that we have simplified here.  A T-P profile
that is coupled with a photochemical model is needed to correctly predict 
some of the feedback mechanisms that occur as photochemistry alters the
atmospheric composition.  Surface processes also much be considered as
they can further affect atmospheric
composition through cycles (such as the carbonate-silicate cycle), outgassing, 
or interactions with a liquid ocean.
Additional questions to be addressed
by a general super-Earth model are those that plague all planetary atmosphere
models such as the presence of clouds and a proper treatment of 
condensation.

Currently, no transiting super-Earths have been detected.  However, 
CoRoT \citep{bag03, bar05} and Kepler 
\citep{bor04, bas05} missions 
are both designed with the
capability of detecting transiting super-Earths orbiting Sun-like stars.  
Depending on occurrence rates, a large number of such planets could be 
discovered in the next 5 years by both of these space-based missions.   
Additionally, ground-based transit surveys such as MEarth \citep{nut07}, which 
will specifically target M-dwarfs, as well
as radial velocity surveys both have the capabilities to detect planets in
the super-Earth range.  Most interestingly, 
the combination of these searches should allow for super-Earths to be
found in a wide
range of orbits and around a variety of types of stars.

\begin{figure}
\includegraphics[angle=90,scale=0.68]{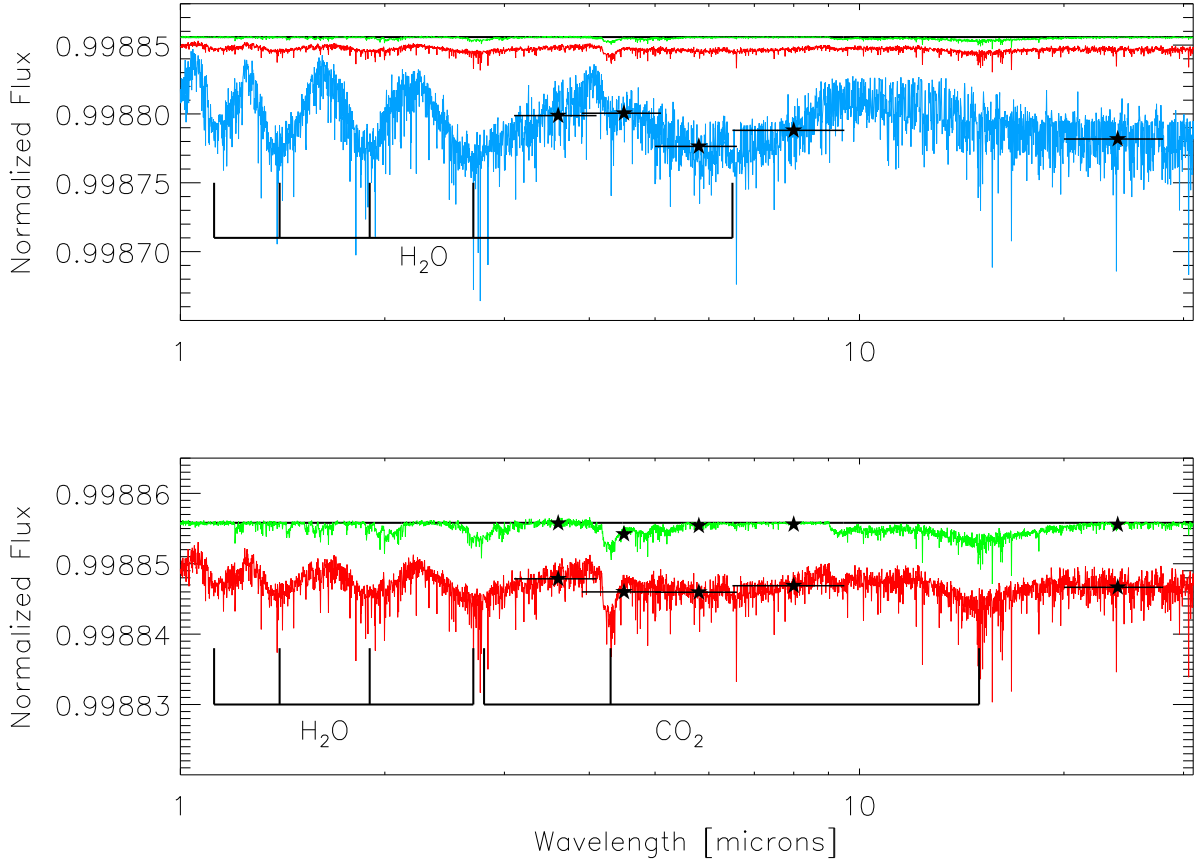}
\caption{Top: Transmission spectra (transit flux divided by out-of-transit 
        flux) of Gl 581c for the cases of a hydrogen-rich 
	atmosphere (blue), an intermediate atmosphere (red), a hydrogen-poor
	atmosphere (green), and no atmosphere (black line).  The stars 
	correspond to the relative flux that would be observed in each of
	the Spitzer IRAC bands as well as in the 24-$\mu$m MIPS band, and the
	attached horizontal bars represent the width of each filter.
	Bottom: Same as above, but zoomed in to show the transmission spectra
	for the intermediate and hydrogen-poor atmospheres only.  Once again, 
	the fluxes expected from the 5 Spitzer bands are overlaid, although the
	horizontal bars have been omitted in the case of the hydrogen-poor 
	atmosphere so as not to obscure the spectrum.  
        \label{transmission}}
\end{figure}

\begin{figure}
\plotone{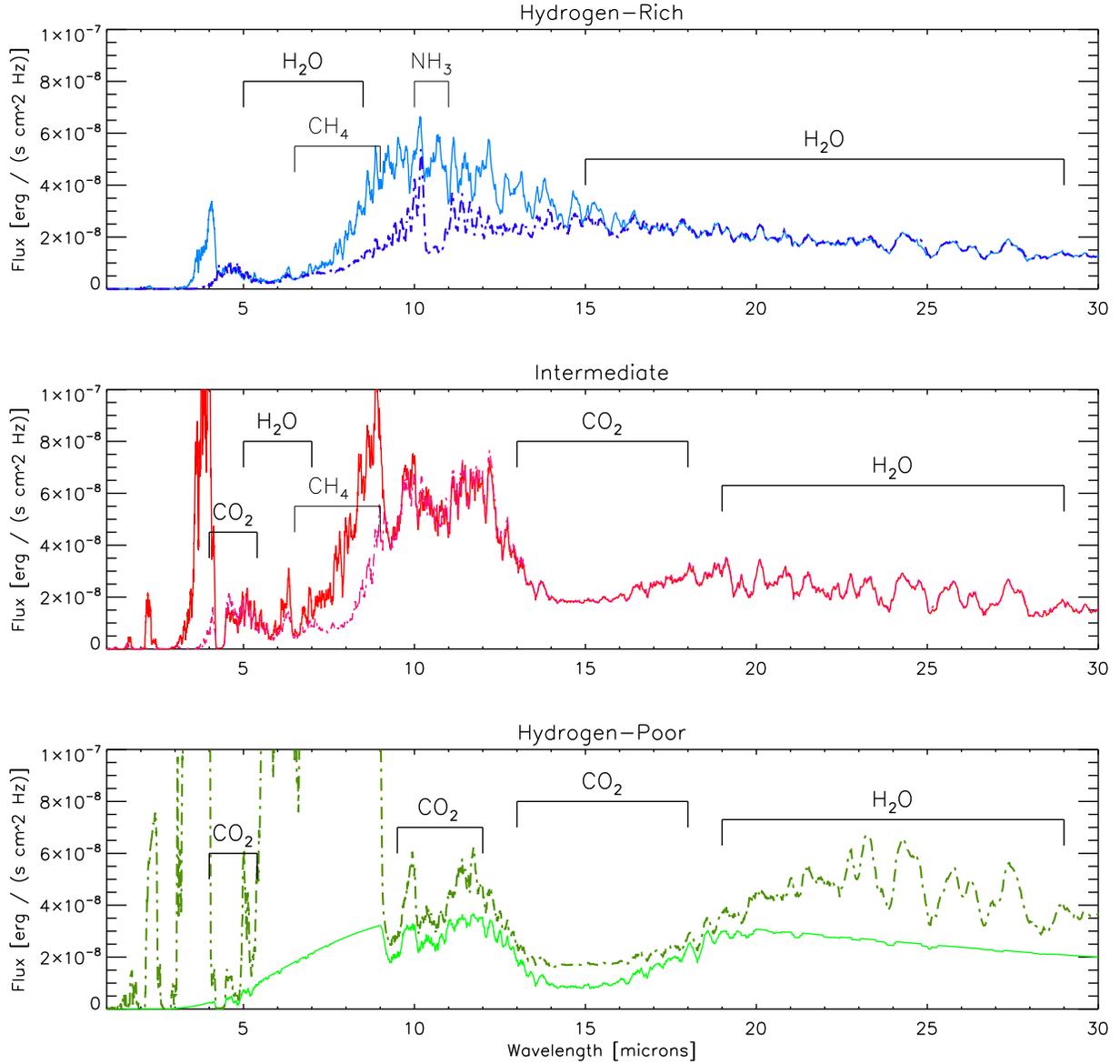}
\caption{Emission spectra of Gl 581c for each of the three cases of atmospheric
        composition considered in this paper.  In the top two panels, the
	solid lines are the spectra with NH$_3$ and CH$_4$ removed due to 
	photochemical considerations.  The dashed lines show the same spectra
	but for an atmosphere in chemical equilibrium.  The bottom panel shows 
	the hydrogen-poor emission spectra for a cloud-free model (dashed line)
	and a model with sulfuric acid clouds at a temperature of 400 K (solid 
	line).  The major features have been labeled, including NH$_3$ and 
	CH$_4$ features in the dashed-line spectra.
        \label{emission}}
\end{figure}

\begin{figure}
\plotone{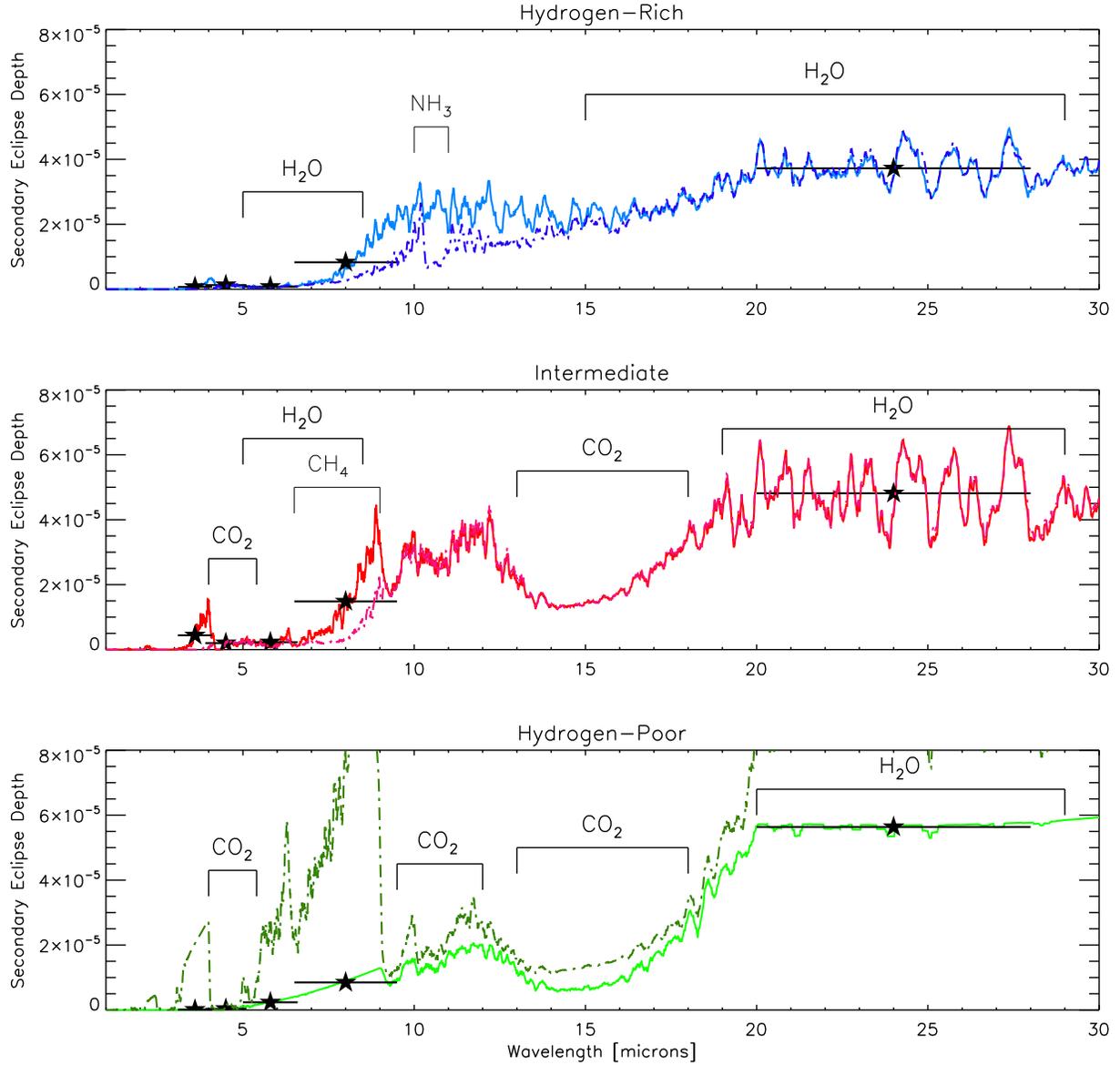}
\caption{Secondary eclipse depths for our three atmosphere scenarios.
        Dashed lines in the top two panels represent models with
	methane and ammonia removed due to photochemical considerations.
	In the bottom panel we include versions of the hydrogen-poor 
	atmosphere both with (solid line) and without (dashed line)
	sulfuric acid clouds.  The major features in the planet's emission 
	spectrum have been labeled.   The model fluxes for the solid-line 
	spectra averaged over the four Spitzer 
	IRAC bands and the 24-$\mu$m  MIPS band are indicated by stars, and
	the solid horizontal lines represent the corresponding filter widths.
        \label{color_mag}}
\end{figure}

Spitzer is currently the only telescope that has demonstrated the capability of
observing exoplanet emission and transmission spectra in the mid-IR.  
Unfortunately, it is unlikely that Spitzer will be able to make the same
types of observations for super-Earths, with the possible exception of a warm
planet with a hydrogen-rich atmosphere -- see Figure ~\ref{transmission}.  
Otherwise, measuements would need to be taken at a level of several to
tens of ppm.  The James Webb Space Telescope (JWST) \citep{gar06} should
fare better, since it has been designed to make these types of measurements at 
the precision needed.  With its
expected instrument complement that will allow for narrow band photometry
and low to medium resolution IR spectroscopy, JWST should allow for
follow-up observations to characterize the atmospheres of transiting 
Super-Earths.  Valenti et al.~(in prep.)~have simulated JWST NIRSpec spectra 
for
transiting Earth-like planets orbiting M-stars, using atmospheric models 
from \citet{ehr06}.
They determine that 10 transits (for a 1.4 hr transit duration) and 28 hours 
of observing time are needed to detect water 
absorption features in transmission at a 4-$\sigma$ detection threshold.
Scaling these values to the hydrogen-rich case for Gl 581c results in
a 20-hour integration time and 6 transits needed to detect its transmission 
features with their expected depth of 10$^{-4}$.  Others have suggested
that JWST can do better by up to a factor of two (Swain 2008, submitted).  If 
this is the case, then even less observing time will be needed.  These types 
of observations
should be feasible with proper telescope scheduling and should allow for the 
full richness of super-Earth atmospheres of different
sizes, compositions, and orbits to be explored.  

\acknowledgements
We would like to thank Richard Freedman for providing some of the molecular
line data for this work, Jonathan Fortney for useful discussion, and our 
anonymous referee for comments that have helped to improve this paper.

\bibliography{ms}

\begin{deluxetable}{llll}
\tablecaption{Atomic Composition\label{tbl1}}
\tablewidth{0pt}
\tablehead{
\colhead{Species} & \colhead{H-Rich} & \colhead{Intermediate} & \colhead{H-Poor}}
\startdata
H & 0.925 & 0.500 & 1.4$\times 10^{-5}$ \\
He & 0.072 & trace & 4.0$\times 10^{-6}$ \\
C & 8.2$\times 10^{-4}$ &  0.128 & 0.325 \\
N & 2.6$\times 10^{-4}$ & 0.040 & 0.024 \\
O & 1.9$\times 10^{-3}$ & 0.304 & 0.651 \\
Ne & 1.3$\times 10^{-4}$ & trace & 2.4$\times 10^{-6}$ \\
S & 2.4$\times 10^{-5}$ & 3.8$\times 10^{-3}$ & 5.1$\times 10^{-5}$ \\

\tablecomments{Abundances are all given as mole fractions.}
\label{composition}
\enddata

\end{deluxetable}

\end{document}